\documentclass[12pt]{article}
\usepackage{a4,epsf}

\def\rms#1{_{\rm #1}}
\def\qcut{\rm q_{cut}}

\newcommand{\bea}{\begin{eqnarray}}
\newcommand{\eea}{\end{eqnarray}}
\newcommand{\simgt}{\hbox{ \raise3pt\hbox to 0pt{$>$}\raise-3pt\hbox{$\sim$} }}
\newcommand{\simlt}{\hbox{ \raise3pt\hbox to 0pt{$<$}\raise-3pt\hbox{$\sim$} }}

\begin{document}
\begin{titlepage}
\title{QCD Potential and $t\bar{t}$ Threshold Cross Section \\
-- Status Report --
}
  \author{M.~Je\.zabek$^{a)}$, J.~H.~K\"uhn$^{b)}$, M.~Peter$^{c)}$,
  Y.~Sumino$^{b)}$\thanks{On leave of absence from Department of Physics,
     Tohoku University, Sendai 980-77, Japan.} \\ and
  T.~Teubner$^{d)}$
  \\ \\ \\ \small
  $a)$ Institute of Nuclear Physics,
       Kawiory 26a, PL-30055 Cracow, Poland \\ \small
  $b)$ Institut f\"ur Theoretische Teilchenphysik,
       Universit\"at Karlsruhe, \\[-5pt] \small
       D-76128 Karlsruhe, Germany \\ \small
  $c)$ Institut f\"ur Theoretische Physik,
       Universit\"at Heidelberg, \\[-5pt] \small
       Philosophenweg 16, D-69120 Heidelberg, Germany \\ \small
  $d)$ Deutsches Elektronen-Synchrotron (DESY), Notkestra\ss e 85, \\[-5pt] 
       \small D-22607 Hamburg, Germany
}
\date{}
\maketitle
\thispagestyle{empty}
\vspace{-5.0truein}
%\author{M.~Je\.zabek$^{a)}$, J.~K\"uhn$^{b)}$, 
%M.~Peter$^{c)}$ and
%Y.~Sumino$^{b)}$\thanks{On 
%leave of absence from Department of Physics, Tohoku University,
%Sendai 980-77, Japan.}
%\\ \\ \\ \small
%$a)$ Institute of Nuclear Physics,
%Kawiory 26a, PL-30055 Cracow, Poland
%\\
%\small
%$b)$ Institut f\"ur Theoretische Teilchenphysik,
%Universit\"at Karlsruhe,\\
%\small
%D-76128 Karlsruhe, Germany
%\\
%\small
%$c)$ Institut f\"ur Theoretische Physik,
%Universit\"at Heidelberg, \\
%\small
%Philosophenweg 16, D-69120 Heidelberg, Germany
%}
\vspace{-7mm}
\begin{flushright}
{\bf DESY\ 98-009}\\
{\bf TTP 98-03}\\
{\bf HD-THEP--98--3}\\
{\bf hep-ph/9801419}\\
{\bf January 1998}
\end{flushright}
\vspace{3.0truein}
\vspace{3cm}
\begin{abstract}
\noindent
{\small
We include the full second-order corrections to the
static QCD potential in the analysis of the $t\bar{t}$
threshold cross section.
Then we examine
the difference between the results obtained in the momentum-space
approach and in the coordinate-space approach, which was
found recently.
Contrary to our expectation, the reduction of the difference
by the inclusion of the second-order corrections is very small.
There still remains a non-negligible deviation, which originates from
the difference in the construction of the potentials in the two spaces.
We scrutinize this problem.
In particular, we estimate our present
theoretical uncertainty of the $t\bar{t}$ threshold cross section 
at the peak to be
$\delta \sigma_{\rm peak}/\sigma_{\rm peak} \simgt 6\%$
within perturbative QCD.
}
\end{abstract}
\vfil

\end{titlepage}
  
In this paper we report on our present theoretical understanding of
the $t\bar{t}$ total cross section near the threshold.
Up to now, all the ${\cal O}(\alpha_s)$ corrections (also 
leading logarithms) have been included in
the calculations of various cross sections near threshold.
In order to take into account the QCD binding effects properly
in the cross sections,
we have to systematically rearrange the perturbative expansion near threshold.
Namely, we first resum all the leading Coulomb singularities 
$\sim (\alpha_s/\beta)^n$, take the result as the leading order contribution, 
and then calculate higher order corrections, which are essentially 
resummations
of the terms
$\sim \alpha_s^{n+1}/\beta^n$, $\alpha_s^{n+2}/\beta^n$, $\ldots$
It is also important to resum large logarithms arising from the large
scale difference involved in the calculation~\cite{sp}.\footnote{
Since the toponium resonance
wave functions have wide distributions $\sim 10$--20~GeV, they probe
a fairly wide range of the QCD potential.
For example,
this is reflected in the fact that 
the fixed-order calculation with any single choice of scale $\mu$
cannot reproduce simultaneously both the distribution and the normalization
of the differential cross section which includes all the leading logarithms.
It is known that
the normalization of the cross section is more 
sensitive to the short-distance behavior of the QCD potential.
}
This is achieved by (first) calculating the Green function of the 
non-relativistic Schr\"{o}dinger equation with the QCD potential~\cite{fk,sp}.
Conventionally 
both the coordinate-space approach developed in Refs.~\cite{sp,sfhmn}
and the momentum-space approach developed in Refs.~\cite{JKT92,JT93}
have been used in solving the equation
by different groups independently.
It has recently been found \cite{ps} that 
there are discrepancies in the results obtained
from the two approaches reflecting the difference in the construction of 
the potentials in both spaces.
It was argued that the differences are formally of ${\cal O}(\alpha_s^2)$ but
their size turns out to be non-negligible.

Quite recently there has been considerable progress in the 
theoretical calculations
of the second-order corrections to the cross section at threshold and
the Coulombic bound-state problem.
New contributions have been calculated analytically~\cite{hoang,hl} and 
numerically~\cite{adkins} for QED bound-states.
Very important steps have been accomplished in QCD as well.
The full second-order correction to the static QCD potential was computed
in~\cite{MP97}.
Also, the ${\cal O}(\alpha_s^2)$ total cross section is known now in the
region $\alpha_s \ll \beta \ll 1$ as a series expansion in $\beta$
\cite{cm}.
All these results have to be included in the calculation of
the full ${\cal O}(\alpha_s^2)$
corrections to the threshold cross section, 
which has been completed just in these days
(as far as the production process of top quarks are 
concerned)~\cite{ht}.

In this report, we incorporate the full ${\cal O}(\alpha_s^3)$ corrections
(the second-order corrections to the leading contribution)
to the static QCD potential 
into our analyses. 
In principle this is a step towards an improvement of the theoretical 
precision in our analysis of the $t\bar{t}$ threshold cross section.
The reduction of the difference is very small, however, 
and there still remains a non-negligible difference.
We scrutinize this problem of the
difference between the momentum-space and the coordinate-space potentials.
We find that there is a theoretical uncertainty within
perturbative QCD which limits our present-day
theoretical accuracy of the threshold cross section.

Let us first state the numerical accuracies attained throughout
our analyses.
We confirmed that our numerical accuracies are at the level of $10^{-4}$.
We have tested our programs with the Coulomb potential whose 
analytical form is known both in momentum space and in coordinate space.
Moreover we confirmed that we obtain the same cross section within the 
above accuracy, irrespective of whether 
we solve the Schr\"odinger equation in momentum space
or first Fourier transform the potential
and solve the Schr\"odinger equation in coordinate space.
In this way we also checked that our numerical Fourier transformation of the
potential (from momentum space to coordinate space) works within the above 
quoted accuracy.
The level of accuracies is quite safe in studying the size of the higher
order corrections which are described in this paper.

Let us now briefly explain the construction of our potentials in momentum space
and in coordinate space, respectively.
More detailed descriptions including formulas are given in the appendices.
The large-momentum part of the momentum-space potential $V_{\rm JKPT}(q)$
is determined as follows.
First the potential has been calculated up to ${\cal O}(\alpha_s^3)$
in a fixed-order calculation.
The result is then improved using the
three-loop renormalization group equation in momentum space.
At low momentum, the potential is continued smoothly to a
Richardson-like potential.
On the other hand, the short-distance part of the
coordinate-space potential $V_{\rm SFHMN}(r)$ 
is calculated by taking the Fourier transform of the fixed-order potential
in momentum space, and
then is improved using the three-loop renormalization group
equation in coordinate space.
At long distance, the potential is continued smoothly
to a phenomenological ansatz.
Thus, it is important to note that the two potentials are {\it not}
the Fourier transforms of each other even in the large-momentum or
short-distance region.
They agree
only up to the next-to-next-to-leading logarithmic terms of 
the series expansion in a fixed $\overline{\rm MS}$ coupling.
The difference begins
with the non-logarithmic term in the three-loop fixed-order correction.

\begin{figure}[htb]\begin{center}
 \epsfxsize 12cm \mbox{\epsfbox{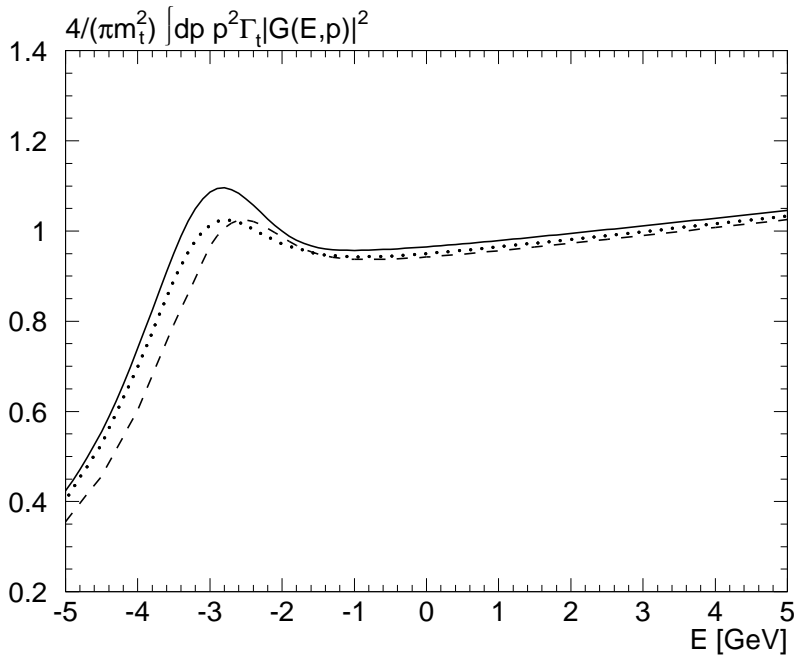}}
 \end{center}
 \caption[]{\label{fig1}Comparison of the total cross sections 
  (normalized to $R$) calculated
  from the different potentials:
  $V_{\rm JKPT}$ (solid), $V_{\rm SFHMN}$ (dashed), and $V_{\rm new}$ (dotted
  line).
  We set $\alpha\rms{\overline{MS}}(M_Z^2) = 0.115$,
  $m_t = 175$~GeV, and $\Gamma_t = 1.427$~GeV.
  }
\end{figure}

In Fig.~\ref{fig1} we show a comparison of the total cross sections 
(normalized to $R$) 
calculated from $V_{\rm JKPT}$ (solid) and from $V_{\rm SFHMN}$ (dashed line),
without any weak or hard-gluon corrections:
\bea
R = \frac{4}{\pi m_t^2}  \int_0^\infty dp \, p^2 |G(E,p)|^2  \, \Gamma_t .
\eea
For the physical parameters we used
$\alpha\rms{\overline{MS}}(M_Z^2) = 0.115$,
$m_t = 175$~GeV, and $\Gamma_t = 1.427$~GeV.
We find that the two cross sections differ by 6.7\% at the 
peaks and by 1.9\% 
at $E = 5$~GeV.\footnote{
In this paper we are not concerned with those differences of the
cross sections which can be absorbed into an additive constant to
the potential, or equivalently, into a redefinition of the top
quark mass.
}
Since the difference of the cross sections calculated from the
next-to-leading order potentials is 8.2\% at the peak and 2.2\%
at $E = 5$~GeV for the same value of $\alpha_s(M_Z^2)$ 
(see Figs.~\ref{scomp} and \ref{compsfhmn}),
the cross sections have come closer only slightly after the inclusion
of the second-order correction to the potential.
The remaining difference is much larger than what one would expect from
an ${\cal O}(\alpha_s^3)$ correction relative to the leading order,
which is not fully included in our analyses, even if we
take into account the high sensitivity to the coupling, 
$\sigma_{\rm peak} \propto \alpha_s^2$\ \cite{sp}.
The purpose of this report is to understand the 
origin of this unexpectedly large difference. 

\begin{figure}
 \begin{center}
 \epsfxsize=12cm \mbox{\epsfbox{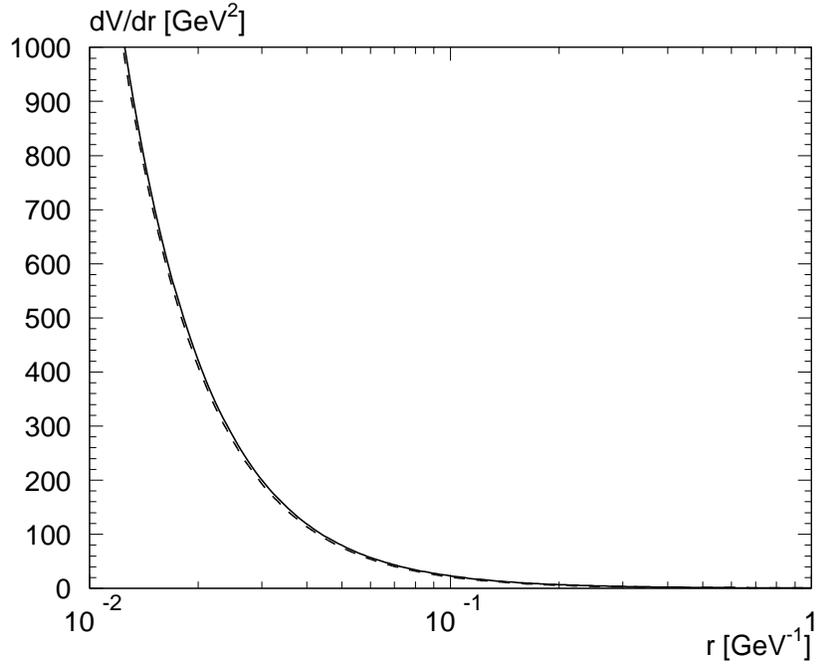}}\\(a)\\
  \epsfxsize=12cm \mbox{\epsffile{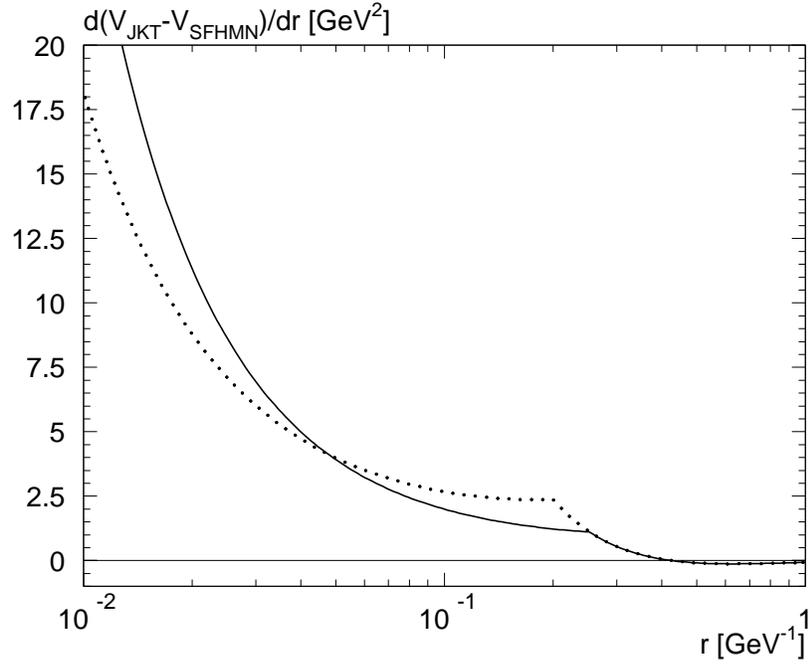}}\\  
  (b)
 \end{center}
 \caption[]{\label{force}
  (a) Comparison of the derivatives of the potentials vs.\ $r$ 
  for $\alpha\rms{\overline{MS}}(M_Z^2) = 0.115$:
  $dV_{\rm JKPT}/dr$ (solid line) and
  $dV_{\rm SFHMN}/dr$ (dashed line).
  (b) Difference of the derivative of the
  potentials vs.\ $r$.
  The solid line shows $\Delta F(r) = dV_{\rm JKPT}/dr - dV_{\rm SFHMN}/dr$,
  and the dotted line shows
  $\Delta F(r) = dV_{\rm JKPT}/dr - dV_{\rm new}/dr$.
  }
\end{figure}

As already mentioned, the difference of the cross sections
reflects the difference of the potentials.
The derivative of the potential $dV(r)/dr$ is directly related to the
size of the cross section;
the cross section is larger if
$dV(r)/dr$ (= magnitude of the attractive force) is larger.  
This is because, with increasing probability that $t$ and $\bar{t}$ 
stay close to each other,
the wave function at the origin $|\psi(0)|^2$ increases, and
so does the total cross section.  
Certainly, adding a constant to $V(r)$ does not affect the size of
the cross section at the peak.
Thus, we Fourier
transformed $V_{\rm JKPT}$ numerically from momentum space to
coordinate space and plot the derivatives of the potentials 
in Fig.~\ref{force}(a).
To demonstrate the difference of the attractive forces, we
show the difference of the derivatives
of the two potentials, 
\bea
\Delta F(r) = \frac{dV_{\rm JKPT}(r)}{dr} - \frac{dV_{\rm SFHMN}(r)}{dr} ,
\eea
(solid line) in Fig.~\ref{force}(b).
We confirm that $\Delta F(r) >0$ holds in the region probed by the
toponium states, $r \sim 0.03$--0.1~GeV$^{-1}$.
One also sees that both potentials have a common slope at 
$r > 0.4$~GeV$^{-1}$ because of the severe constraints from
the bottomonium and charmonium data.
The kink seen in the figure is due to a discontinuity of 
$d^2 V_{\rm SFHMN}/dr^2$ located at the continuation point, $r = r_c$.

\begin{figure}
\begin{center}
  \leavevmode
  \epsfxsize=12cm 
  \mbox{\epsffile{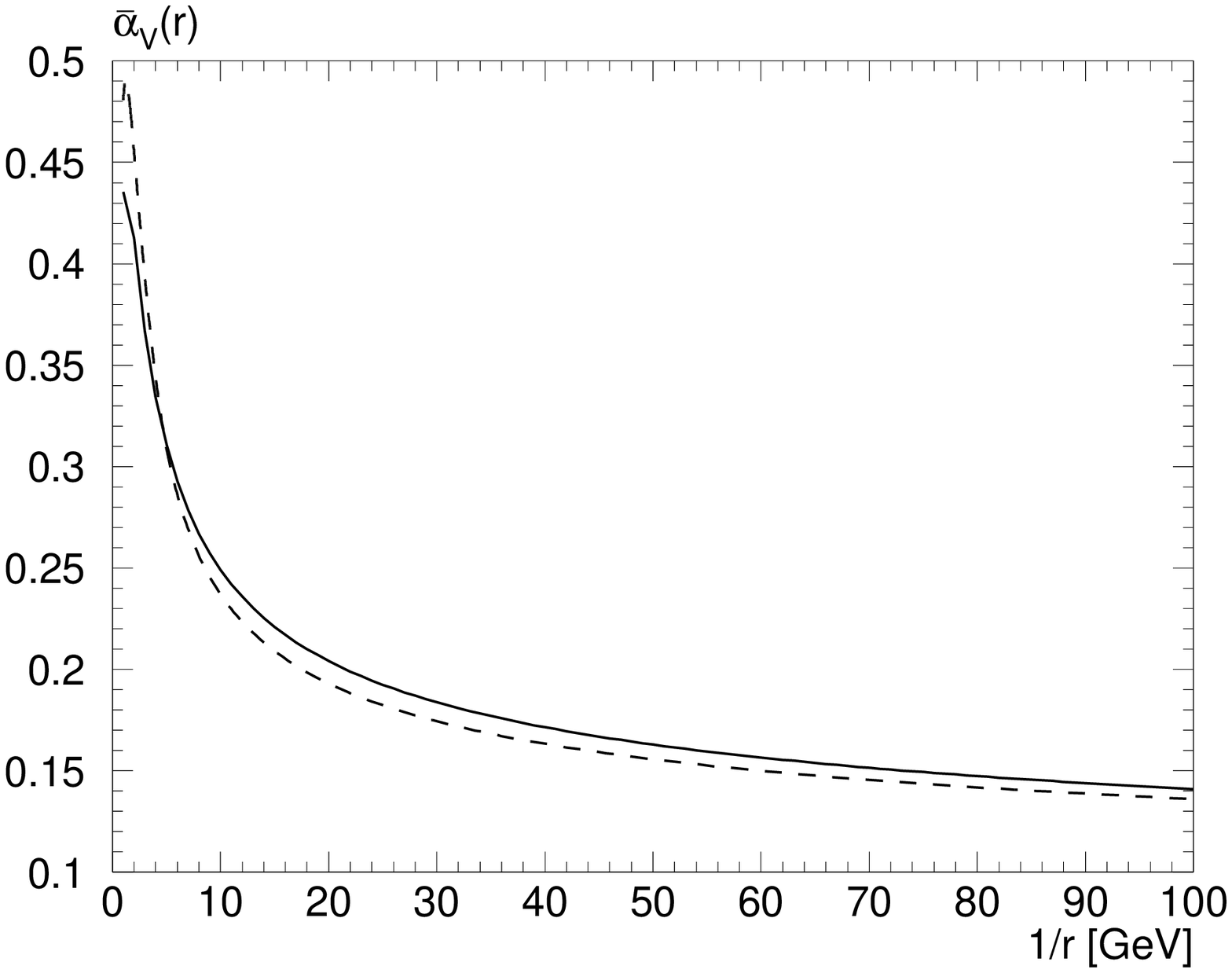}}\\(a)\\
  \epsfxsize=12cm \mbox{\epsffile{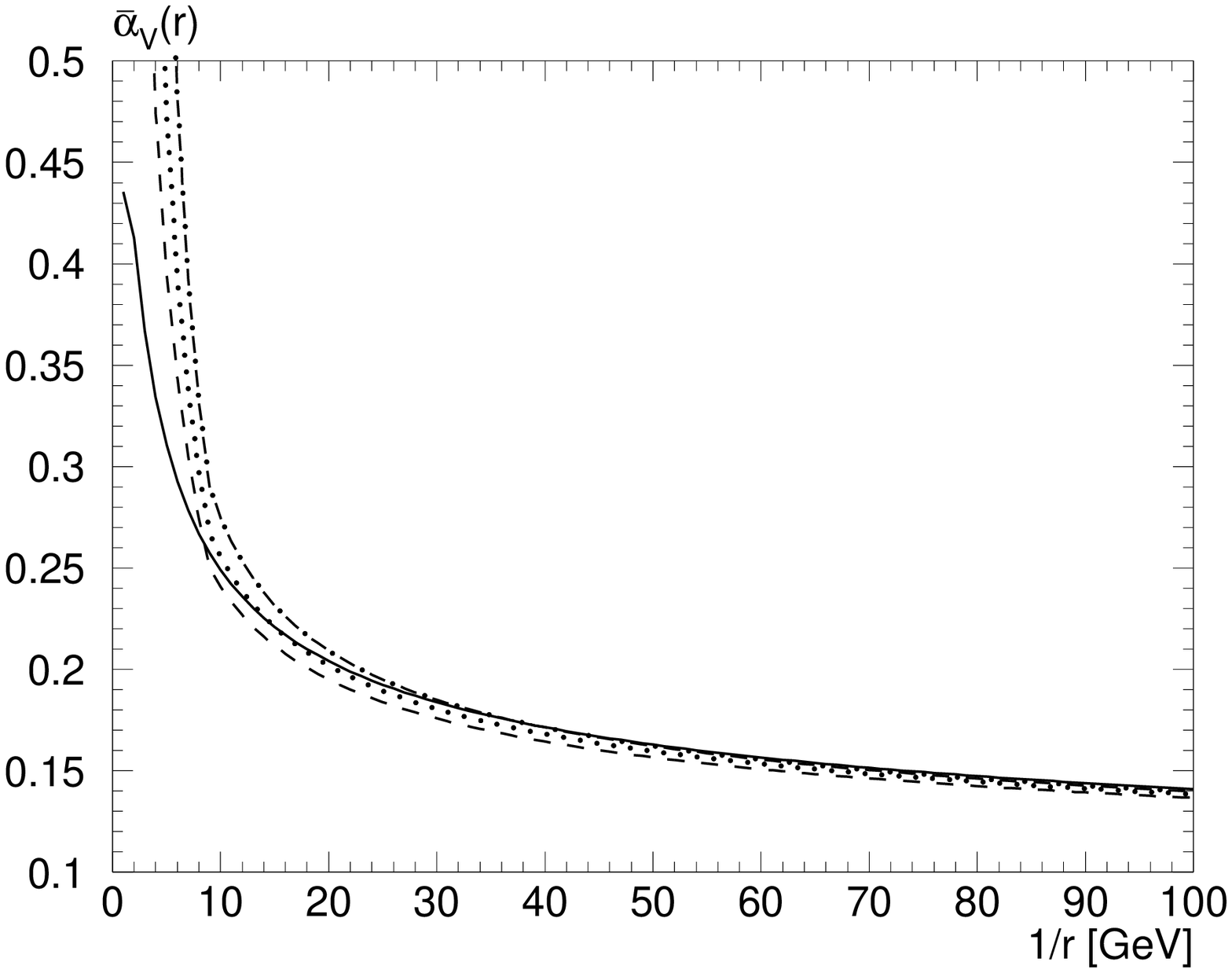}}\\  
  (b)
\end{center}
  \caption[]{
        \label{eff}
(a) Comparison of the coordinate-space effective charges defined
    from $V_{\rm JKPT}$ (solid) and from $V_{\rm SFHMN}$ (dashed line).
(b) Comparison of the coordinate-space effective charges defined
    from the various terms of Eq.~(\ref{rel}).
    See the text for the description of each curve.
}
\end{figure}

In order to compare the asymptotic behavior of the potentials 
more clearly,
we plot in Fig.~\ref{eff}(a) the coordinate-space effective couplings 
defined by
\bea
\bar{\alpha}_{\rm V} ( 1/r ) 
= \left( - C_F / r \right)^{-1} V(r) 
\label{cooreff}
\eea
for $V_{\rm JKPT}(r)$ and $V_{\rm SFHMN}(r)$ as
solid and dashed lines, respectively.
Contrary to our expectation,
the difference of the couplings exceeds 3\% even at very 
short distances, $1/r \simeq 100$~GeV.

Naturally the question arises: Why is there such a large discrepancy
between the potential constructed in momentum space and that
constructed in coordinate space?
To answer this question, let us examine a
relation connecting the effective coupling
in coordinate space, defined by Eq.~(\ref{cooreff}),
and the effective coupling in 
momentum space, defined from the momentum-space potential as
\bea
\alpha_{\rm V} ( q ) 
= \left( - 4 \pi C_F / q^2 \right)^{-1} V(q) .
\eea
The relation is derived from the renormalization group equation 
of $\alpha_{\rm V}(q)$ and 
exact to all orders. 
In
the asymptotic region where the couplings are small, 
it can be given in the form of an asymptotic series~\cite{jps},
which reads numerically
\bea
\bar{\alpha}_{\rm V}(1/r) = 
\alpha_{\rm V} + 1.225 \, \alpha_{\rm V}^3 
+ 5.596 \, \alpha_{\rm V}^4
+ 32.202 \, \alpha_{\rm V}^5 + \ldots
\label{rel}
\eea
for $n_f = 5$.
On the right-hand-side, 
$\alpha_{\rm V} = \alpha_{\rm V}(q = e^{-\gamma_E}/r)$.
All terms which are
written explicitly are determined from the known coefficients 
of the $\beta$ function, $\beta_0^V$, $\beta_1^V$, and $\beta_2^V$.
At present, we can use the above relation consistently only at 
$O(\alpha_{\rm V}^3)$ because we know the effective couplings
only up to the next-to-next-to-leading order corrections in
perturbative QCD,
i.e. we know the relation between 
$\alpha_{\rm V}$ and $\alpha\rms{\overline{MS}}$ only up to
${\cal O}(\alpha\rms{\overline{MS}}^3)$.
Due to this limitation, essentially, the effective coupling 
$\bar{\alpha}_{\rm V}$ defined from $V_{\rm SFHMN}$ 
is the right-hand-side of the above
equation truncated at the $O(\alpha_{\rm V}^3)$ term, 
while $\bar{\alpha}_{\rm V}$ defined from $V_{\rm JKPT}$ 
is the right-hand-side including all terms.
Numerically, the $O(\alpha_{\rm V}^4)$ term and the
$O(\alpha_{\rm V}^5)$ term
contribute as +1.5\% and +1.2\% corrections, respectively, for 
$\alpha_{\rm V} \simeq 0.14$ (corresponding to $1/r = 100$~GeV).
Therefore, these higher order
terms indeed explain the difference of the effective couplings
at small $r$.
Fig.~\ref{eff}(b) shows several curves derived from the above relation:
\begin{enumerate}
\item
The solid line is $\bar{\alpha}_{\rm V}(1/r)$ defined from
$V_{\rm JKPT}$.
\item
The dashed curve is 
$\alpha_{\rm V} + 1.225 \, \alpha_{\rm V}^3$, where
$\alpha_{\rm V} = \alpha_{\rm V}(q = e^{-\gamma_E}/r)$
is calculated using the perturbative prediction in momentum space.
This curve is essentially the same as 
$\bar{\alpha}_{\rm V}(1/r)$ defined from
$V_{\rm SFHMN}(r)$, since it is the next-to-next-to-leading order 
perturbative prediction for the coordinate-space coupling at
short distances.
\item
The dotted curve includes the next correction, 
$5.596 \, \alpha_{\rm V}^4$, which is in fact even larger than the 
${\cal O}(\alpha_{\rm V}^3)$ term below $1/r \sim 30$~GeV.
\item
The dash-dotted curve includes the ${\cal O}(\alpha_{\rm V}^5)$ term.
\end{enumerate}
We observe that the agreement of both sides of 
Eq.~(\ref{rel}) becomes better as
we include more terms at small $r$, while it becomes worse at large $r$ on
account of the asymptoticness of the series.
From the purely perturbative point of view, the discrepancy between our 
two potentials, $V_{\rm JKPT}$ and $V_{\rm SFHMN}$, in the asymptotic
region thus 
seems real, an indication of large higher order corrections.\footnote{
If we apply the same method (the relation between $\bar{\alpha}_{\rm V}$ 
and $\alpha_{\rm V}$) to estimate the size of the already known 
${\cal O}(\alpha_s^3)$ correction, we obtain $\pi^2 \beta_0^2/3 = 193.4$,
which turns out to be a slight under-estimate of the true correction
$a_2 = 333.5$~\cite{MP97} ($n_f=5$).
}
If the third-order correction to the potential will ever be computed
in the future, the ${\cal O}(\alpha_V^4)$ term will be treated consistently
and the difference will reduce by 1.5\% at $1/r \simeq 100$~GeV.

The above 3\% uncertainty of $\bar{\alpha}_{\rm V}(1/r)$ at 
$1/r \simeq 100$~GeV
provides a certain criterion for the present theoretical uncertainty
of the $t\bar{t}$ cross section.
In fact, it would already limit the theoretical accuracy of 
$\bar{\alpha}_{\rm V}(1/r)$ at longer distances to be
not better than 3\%.
If we combine this with a naive estimate
$\sigma_{\rm peak} \propto \bar{\alpha}_{\rm V}^2$,
we expect a theoretical uncertainty of the peak cross section to be
$\delta \sigma_{\rm peak}/\sigma_{\rm peak} \simgt 6\%$.
Therefore, the large discrepancy of the cross sections which we have
seen turns out to be quite consistent with this estimated uncertainty.

One is tempted to include one more term of the above series~(\ref{rel})
to define a (new) coordinate-space potential
despite our ignorance of the corresponding terms in the relation
between $\alpha_{\rm V}$ and $\alpha\rms{\overline{MS}}$, since 
this would apparently reduce the difference between the two 
effective couplings. 
In fact we did this exercise, but (to our surprise) it did not bring
the cross section closer to the one calculated
from the momentum-space potential $V_{\rm JKPT}$.  
This cross section calculated from the potential $V_{\rm new}(r)$,
which incorporates the ${\cal O}(\alpha_{\rm V}^4)$ term
of Eq.~(\ref{rel}),
is shown as a dotted curve in Fig.~\ref{fig1}(b).\footnote{
The shift of the peak position to lower energy is caused mostly by 
a decrease of the constant $c_0$ in Eq.~(\ref{vsfhmn}) and
not due to an increase of the attractive force.
Since the effective coupling $\bar{\alpha}_{\rm V}(1/r)$ runs faster
for $V_{\rm new}$,
the perturbative potential is connected to the intermediate-distance
phenomenological potential at a deeper point.
}

We may understand the reason why the cross section did not
approach that of $V_{\rm JKPT}$ if
we look at the difference of the ``forces'',
$\Delta F(r) = dV_{\rm JKPT}/dr - dV_{\rm new}/dr$, 
shown as a dotted line in Fig.~\ref{force}(b):
it can be seen that, upon inclusion of the ${\cal O}(\alpha_{\rm V}^4)$ term,
the difference $\Delta F(r)$ decreased at small distances,
$r < 0.05$~GeV$^{-1}$, as expected, whereas $\Delta F(r)$ {\it increased}
at distances $r > 0.05$~GeV$^{-1}$ which is still in the range
probed by the toponium states.
It is due to a compensation between the decrease and increase of $\Delta F(r)$
that the normalization of the cross section scarcely changed.
The increase of $\Delta F(r)$ at large distances
results from the bad convergence of the asymptotic series,
Eq.~(\ref{rel}), for a large coupling, as we have already seen in
Fig.~\ref{eff}(b).
This fact indicates that we are no longer able to improve the agreement
of the cross sections by including even higher order terms,
as we are confronting the problem of asymptoticness of the series.

Some indications can be obtained by looking into the nature of 
the perturbative expansion of each potential.
Within our present knowledge of the static QCD potential,
the perturbative series looks more convergent
for the momentum-space potential than for the coordinate-space
potential.
To see this, one may compare 
the $\beta$ functions of the effective
couplings (the V-scheme couplings) in both spaces~\cite{MP97}.
Numerically, the first three terms in the perturbative expansion 
read
\begin{itemize}
\item
(momentum-space coupling)
\bea
&&\mu^2 \frac{d \alpha_{\rm V}}{d \mu^2}
= - 0.6101 \, \alpha_{\rm V}^2
    - 0.2449 \, \alpha_{\rm V}^3
    - 1.198 \, \alpha_{\rm V}^4 + \ldots 
\eea
\item
(coordinate-space coupling)
\bea
&&\mu^2 \frac{d \bar{\alpha}_{\rm V}}{d \mu^2}
= - 0.6101 \, \bar{\alpha}_{\rm V}^2
    - 0.2449 \, \bar{\alpha}_{\rm V}^3
    - 1.945 \, \bar{\alpha}_{\rm V}^4 + \ldots 
\eea
\end{itemize}
for $n_f = 5$.
The first two coefficients are universal.  
The third coefficient depends on the scheme (the definition) of
the coupling.
As the third coefficients for the V-scheme couplings are quite
large, the third term of the $\beta$ function is comparable to
the second term already for $\alpha_V = 0.20$ and for
$\bar{\alpha}_V=0.13$, respectively.\footnote{
This is the reason why we evolve the $\overline{\rm MS}$ coupling 
instead of evolving the V-scheme couplings using their
own $\beta$ functions.
Otherwise we would have lost the reasoning to
keep the third term of the $\beta$ function at a fairly
large momentum/short distance.
For comparison, the $\beta$ function of the
$\overline{\rm MS}$ coupling for the same $n_f$ is given by
$$
\mu^2 \frac{d \alpha\rms{\overline{MS}}}{d \mu^2}
= - 0.6101 \, \alpha\rms{\overline{MS}}^2
  - 0.2449 \, \alpha\rms{\overline{MS}}^3
  - 0.09116 \, \alpha\rms{\overline{MS}}^4 + \ldots .
$$
}
The difference of the third coefficients between momentum space
and coordinate space originates from the
$\pi^2\beta_0^2/3$ term in Eq.~(\ref{coorptpot}),
which comes from the Fourier transformation.
(Compare Eqs.~(\ref{pteff}) and (\ref{coorptpot}).)
Although the magnitude of the third coefficients is of the same order, 
in practice it makes a certain difference whether an apparent 
convergence is lost at
$\alpha_V = 0.20$ or $\bar{\alpha}_V = 0.13$
because there is a large scale difference between the two values.
This indicates a worse convergence in coordinate space than
in momentum space.

If we evolve the coordinate-space coupling
$\bar{\alpha}_{\rm V}$ using its own $\beta$ function
up to the third term, the coupling exhibits an infrared
pole at $1/r = \Lambda \sim 2$~GeV, which
is an order of magnitude larger than $\Lambda_{\overline{\rm MS}}$
of the $\overline{\rm MS}$ coupling.
The asymptoticness of the series in Eq.~(\ref{rel}) is 
closely related to the existence of this pole.
In fact, one may estimate the uncertainty caused by the
asymptoticness of the expansion to be 
$\delta \bar{\alpha}_{\rm V}(1/r) \sim 
\Lambda r + (\Lambda r)^2 + \ldots$\ \cite{jps}.
If we translate this to the uncertainty in
the slope of the coordinate-space
potential, we obtain $\delta F(r) \sim \Lambda^2$.
This is in good agreement with the discrepancy 
$\Delta F(r) \sim 1$--4~GeV$^2$ 
in the region $r > 0.05$~GeV$^{-1}$ (see Fig.~\ref{force}), where
the usability of the asymptotic expansion is already limited to the 
first two or three terms.
(For $r < 0.05$~GeV$^{-1}$, one may reduce the difference by including
more terms.)

It is interesting to examine the level of uncertainties within
each of the momentum-space approach and the coordinate-space 
approach by itself.
\begin{figure}
 \begin{center}
 \epsfxsize=12cm \mbox{\epsfbox{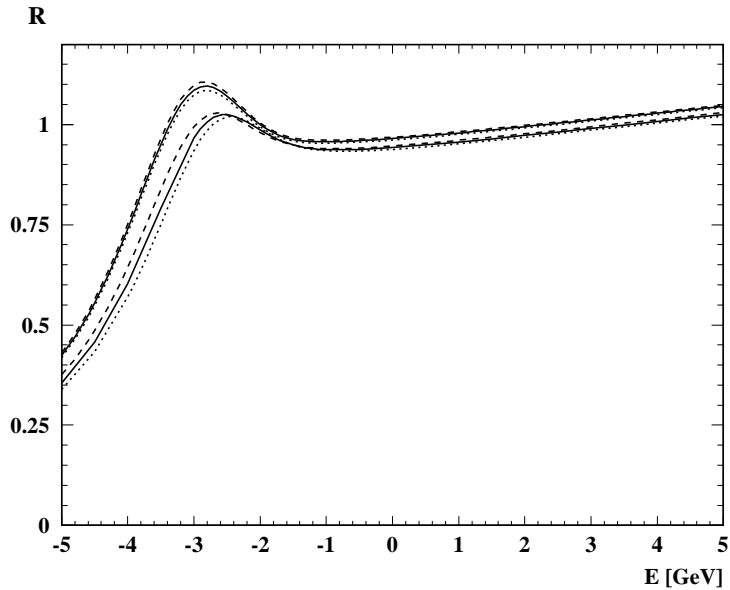}}
 \end{center}
 \caption[]{\label{scalevar}
Comparison of the cross sections for different choices of the scale.
The upper three curves are for the momentum-space approach:
$\mu = q$ (solid), $\mu = \sqrt{2}q$ (dotted), and $\mu = q/\sqrt{2}$
(dashed line).
The lower three curve are for the coordinate-space approach:
$\mu = \mu_2 = {\rm exp}(-\gamma_E)/r$ (solid), 
$\mu = \sqrt{2}\mu_2$ (dotted), 
and $\mu = \mu_2/\sqrt{2}$
(dashed line).
}
\end{figure}
Fig.~\ref{scalevar}
shows how the cross section changes when we vary the 
scale by a factor of 2 in each approach: from
$\mu = q/\sqrt{2}$ to $\mu = \sqrt{2}q$ in Eq.~(28) of \cite{MP97} in
the momentum-space approach (upper three curves), and
from $\mu = \mu_2/\sqrt{2}$ to 
$\mu =\sqrt{2}\mu_2$ in Eq.~(44) of \cite{MP97} in the
coordinate-space approach (lower three curves).
For the momentum-space approach,
the variation of the cross section is about 2\% at the peak
and around 0.6\% for c.m.\ energies above threshold.
Meanwhile in the coordinate-space approach,
the variation of the cross section amounts to 0.9\% at the peak
and 0.8\% at larger c.m.\ energies.
These results may be regarded as an internal consistency check for each
approach and even
as a sign for the stability of the theoretical predictions.
Nevertheless one should keep in mind that 
the internal consistency is not the same as the accuracy of the 
theoretical predictions.
Since the ${\cal O}(\alpha_s^2)$
corrections to the potential resulted in an unexpectedly large modification
of the total cross section (see Figs.~\ref{scomp} and \ref{compsfhmn}),
it would be legitimate to consider each of
our results as accurate only if the same method could estimate the size
of the next-to-next-to-leading order correction reasonably well and hence if the 
cross section became considerably less sensitive to the scale variation
after including this correction.
This is not the case in our problem, however.
The very existence of a large constant at the next-to-next-to leading
order ($a_2$ in~\cite{MP97}, see also Eqs.~(\ref{pteff}) and 
(\ref{coorptpot})), which generates these large modifications,
may indicate also large corrections at even higher orders.
Furthermore, if we consider the full set of 
the fixed-order ${\cal O}(\alpha_s^2)$ corrections to the cross section near
threshold~\cite{ht}, they are larger in size and even more scale
dependent than the corrections to the potential alone.
The theoretical uncertainty may therefore be larger than indicated by the
study in this paper.

Still there may be some possibilities to reduce the difference between
the momentum-space potential and the coordinate-space potential in
the region probed by the toponium states.
An obvious point to be improved is to remove the discontinuity
of $V_{\rm SFHMN}''(r)$ at $r = r_c$.
If we employed a smoother interpolation of the 
perturbative potential to the
intermediate-distance potential, we would have a better
agreement of the two cross sections, see Fig.~\ref{force}(b).
This tendency is expected due to the specific interpolation
method adopted for $V_{\rm SFHMN}(r)$.\footnote{
$V_{\rm SFHMN}(r)$ matches the perturbative potential exactly up to
a vicinity of the infrared pole.
Since the rapid acceleration of the running of 
$\bar{\alpha}_{\rm V}(1/r)$ towards
the pole tends to amplify the deviation from $V_{\rm JKPT}(r)$,
if we employed an analytic 
regularization of the pole while keeping the potential to 
approximate the perturbative potential at short distances,
the running of $\bar{\alpha}_{\rm V}(1/r)$ should be tamed, and hence 
the potential should come closer to $V_{\rm JKPT}(r)$.
}
However, one has to be careful with this argument.
The slope of the potential in the intermediate-distance
region is fixed by experimental data, which
correspond to one fixed value of 
$\alpha\rms{\overline{MS}}(M_Z^2)$ (= the true value in nature).
We are interpolating the prediction of perturbative QCD,
which obviously depends on our input value
of $\alpha\rms{\overline{MS}}(M_Z^2)$,
to a phenomenological potential, which is independent of 
it.
This means, we do expect a non-smooth transition for any
value of $\alpha\rms{\overline{MS}}(M_Z^2)$ different from the
true value.
Moreover, if we want to extract the value of 
$\alpha\rms{\overline{MS}}(M_Z^2)$ by comparing the theoretical
predictions to the experimentally measured cross section, the sensitivity
to $\alpha\rms{\overline{MS}}(M_Z^2)$ decreases if the predictions 
depend on the way we perform the interpolation.
Ideally we would want to have an intermediate-distance potential as the 
prediction of QCD --- necessarily non-perturbative --- for a given 
input value of 
$\alpha\rms{\overline{MS}}(M_Z^2)$.

It would be important to understand the problem of the difference
in the potentials also in momentum space, at 
least as much as we do in coordinate space presently.  
We have not done this analysis so far because of the difficulty
in the numerical Fourier transformation of the potential
from coordinate space to momentum space.

\section*{\bf Summary}
\begin{itemize}
\item
There is a difference between the 
potential constructed
in momentum space and that constructed in coordinate space even at
a fairly short-distance, $1/r \sim 100$~GeV.
The difference can be understood within the framework of perturbative QCD.
We already know that there is a large correction at ${\cal O}(\alpha_s^4)$
in the relation between the two potentials, although
a consistent treatment is not possible
until the full ${\cal O}(\alpha_s^4)$ corrections to the
QCD potential are calculated.

\item
The above difference at short distances
provides a criterion for our present theoretical accuracy
of the $t\bar{t}$ cross section, 
$\delta \sigma_{\rm peak}/\sigma_{\rm peak} \simgt 6\%$.

\item
In addition, it seems that
we are confronting the problem of the asymptoticness of the
perturbative series in the calculation of 
the $t\bar{t}$ cross section, as the 
top quarks do not probe a region which is sufficiently deep
in the potential.
We may not be able to improve our 
theoretical precision
even if the higher order corrections are calculated in perturbative QCD.

\item
We may, however, discuss which of the two approaches gives a more
favorable result theoretically.
Up to the second-order corrections, the perturbative series looks 
better convergent
for the momentum-space potential than for the coordinate-space potential.

\end{itemize}

\section*{\bf Acknowledgements}
This work is partly supported by BMBF grant POL-239-96, by
KBN grant 2P03B08414 and by the Alexander von Humboldt Foundation.

% beta0, beta1, beta2   7.66666666666667        38.6666666666667     
%   180.907407407408     
% beta2V, beta2V_coor   2377.95386720642        3860.46592582438     
%  0.610093948518932       0.244859527135650       9.116471061282584E-002
%   1.19832282857450        1.94540546460810     
%   333.495056216404        193.371138080603 

\begin{appendix}
\section{The momentum-space potential}

It seems appropriate to describe the potential $V_{\rm JKPT}$ used in
the present analysis in some more detail. 
This potential is very similar to the potential $V_{\rm JKT}$ described
in~\cite{JT93} and used in all later numerical studies within 
the momentum-space framework.
It includes, however, the next-to-next-to-leading order terms 
from~\cite{MP97}.
The momentum-space potential can be written as
\begin{equation}
  V_{\rm JKPT}({\bf q}) = V_0(\qcut)\cdot(2\pi)^3\delta({\bf q}) - 4\pi C_F
    \frac{ \alpha_{\rm JKPT}({\bf q}^2) }{ {\bf q}^2 }.
\label{VJKPT}
\end{equation}
The effective coupling $\alpha_{\rm JKPT}$ is defined to coincide with
the two-loop perturbative prediction for large momenta, to be
Richardson-like for small momenta, and to simply interpolate between
these two shapes in some intermediate range:
\begin{equation}
  \alpha_{\rm JKPT}({\bf q}^2) = \left\{ 
  \begin{array}{ll}
    \alpha\rms{V,pert}({\bf q}^2), & |{\bf q}|> {\rm q}_1=5\ \mbox{GeV} \\
    \alpha\rms{Rich}({\bf q}^2),   & |{\bf q}|< {\rm q}_2=2\ \mbox{GeV} \\
    \alpha\rms{Rich}({\bf q}^2) + \frac{|{\bf q}|-{\rm q}_2}{\rm q_1-q_2}\Big(
     \alpha\rms{V,pert}({\rm q}_1^2)-\alpha\rms{Rich}({\rm q}_1^2)\Big),
     & {\rm q}_2 < |{\bf q}| < {\rm q}_1.
  \end{array}\right.
\label{momeff}
\end{equation}
The intermediate regime is only introduced to obtain a smoother transition
between the small and large momentum parts, respectively.

The first difference between the updated potential and the former version
is the fact that we are now able to use the full two-loop expression for the
perturbative part,
\begin{equation}
  \alpha\rms{V,pert}({\bf q}^2) =
  \alpha\rms{\overline{MS}}({\bf q}^2)\Bigg( 1 +
  a_1 \frac{\alpha\rms{\overline{MS}}({\bf q}^2)}{4\pi} +
  a_2 \Big(\frac{\alpha\rms{\overline{MS}}({\bf q}^2)}{4\pi}\Big)^2\Bigg),
\label{pteff}
\end{equation}
with the coefficients $a_1$ and $a_2$ given in~\cite{MP97}. As the $b$-quark
threshold is neglected, $n_f=5$ is set throughout in the evolution of
the $\overline{\rm MS}$-coupling, which can now consistently be performed
at three-loop accuracy.

A Richardson-like behavior for small momenta is chosen
since the Richardson potential~\cite{Rich79} is known to describe
the charmonium and bottomonium spectra fairly well. A pure Richardson form,
however, would lead to severe numerical problems. Hence the ansatz has to
be modified slightly by introducing two ``subtraction terms'',
\begin{equation}
  \alpha\rms{Rich}({\bf q}^2) =
  \frac{4\pi}{\beta_0(n_f=3)}\Bigg( \frac{1}{
  \ln(1+\frac{{\bf q}^2}{\Lambda\rms{R}^2})}-\frac{\Lambda\rms{R}^2}{{\bf q}^2}
   + \frac{\Lambda\rms{R}^2}{{{\bf q}^2}+\qcut^2} \Bigg)
\end{equation}
with $\Lambda\rms{R}=400$~MeV.
The first subtraction regulates the divergent behavior for $|{\bf
q}| \to0$, the second subtraction is designed to reduce the modification
introduced through the first to a minimum. Without the second additional term,
the linear part of the position-space Richardson potential would be
removed completely, whereas with it the first subtraction is cancelled
for ${\bf q}^2\gg \qcut^2$ , and thus a big part
of the confining potential is kept. It thus seems desirable to choose the
parameter $\qcut$ small, but
evidently it cannot be put to zero to really recover the pure Richardson
potential. 
However, the linear part of the potential plays practically no role for 
the $t\bar t$-system as will be demonstrated below. The exact value
of $\qcut$ is therefore relatively unimportant and
the adopted value $\qcut=50$~MeV results in both
numerical efficiency and speed of the program and 
a fairly good accuracy of the predictions.

The constant  $V_0(\qcut)$ in Eq.~(\ref{VJKPT}) is to some extent an
arbitrary parameter. Different choices of  $V_0(\qcut)$ reflect the
ambiguity in the definition of the pole masses for confined quarks.
For $V_{\rm JKPT}({\bf q})$ the choice
\begin{equation}
  V_0(\qcut) = \frac{4\pi C_F}{\beta_0(n_f=3)}
       \frac{\Lambda\rms{R}^2}{\qcut}
\end{equation}
is used. It leads to a Richardson-like potential that depends only
weakly on the parameter $\qcut$ and coincides with the true
Richardson potential in position space in the limit $\qcut\to0$.
With this potential one obtains for the pole mass of the $b$ quark
$m_b=4.84$~GeV.
The choice of $V_0$ is the
second difference to the potential used in earlier works, where the constant
$V_0$ was fixed by the condition $V_{JKT}(r=1$\ GeV$^{-1})=-1/4$~GeV leading
to $m_b=4.7$~GeV.  

\begin{figure}[htb]\begin{center}
 \epsfxsize 12cm \mbox{\epsfbox{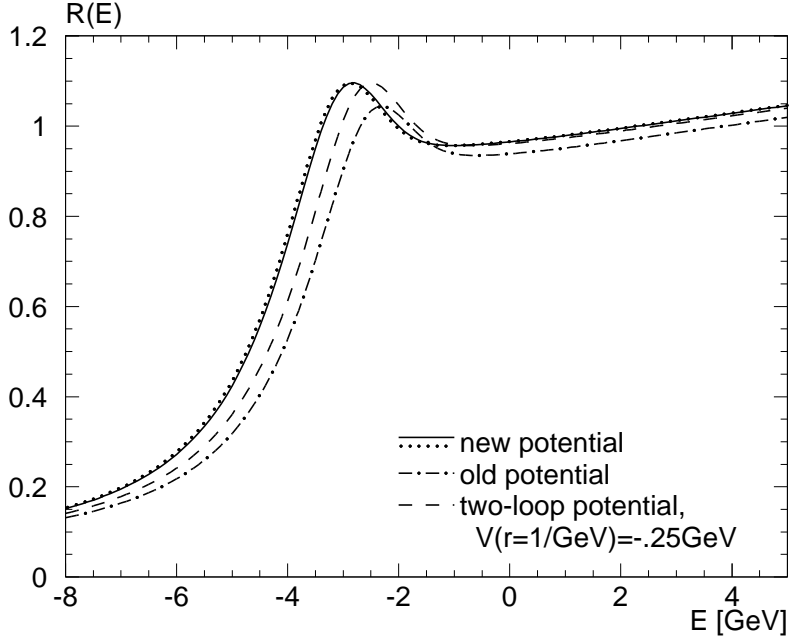}}
 \end{center}
 \caption[]{\label{scomp}Comparison of the total cross section 
  (normalized to $R$) for
  $t\bar t$-production as a function of $E=\sqrt{s}-2m_t$ for $m_t=175$~GeV,
  $\alpha\rms{\overline{MS}}=0.115$, and $\Gamma_t = 1.427$~GeV,
  using the different potentials described in the text. The dash-dotted line
  corresponds to the model of older analyses, the dashed line to the inclusion
  of the two-loop correction to the perturbative part. The solid line shows
  the prediction of the new model, which differs from the dashed line only
  by a different definition of the energy. The dotted line corresponds to the
  new potential with the linear part in position space removed.}
\end{figure}

Fig.~\ref{scomp} shows a comparison of the total cross section for
$t\bar t$-production near threshold as a function of energy using the
``old'' potential, i.e.\ the one as described but
using the one-loop formula for $\alpha\rms{V,pert}$ (and correspondingly the
two-loop evolution for the $\overline{\rm MS}$-coupling) and fixing
$V_0$ through $V(r=1$\ GeV$^{-1})=-1/4$\ GeV (dash-dotted line), 
and the ``new''
potential (solid line). There are two changes: first,
the inclusion of the two-loop correction to the perturbative potential
increases the strength of the attractive interaction between $t$ and $\bar t$,
and thus leads to an increase in
the cross section. This is nicely demonstrated by the dashed curve, which
corresponds to the inclusion of the two-loop potential, but the old
choice of $V_0$. Second, the modified choice for $V_0$ leads
to a small shift of about 300 MeV in the energy scale, which is just the
difference between the two $V_0$. The dotted curve has been included to
once again demonstrate that the $t\bar t$ system is quite insensitve
to the long-range part of the potential: this curve corresponds to
the choice $\qcut=\infty$, i.e.\ to completely removing the linear
part from the potential and setting $V_0=0$.

\section{The coordinate-space potential}

The short-distance part of the coordinate-space potential is
given by the next-to-next-to-leading order static QCD potential
in position space \cite{MP97}, 
whereas its form in the intermediate- and long-distance
region is determined phenomenologically.
We thus have
\bea
V_{\rm SFHMN}(r) = \left\{
\begin{array}{ll}
V_{\rm pert}(r)
& {\rm at}~~ r<r_c\,, \\
c_0 + c_1 \log (r/r_0) \exp (-r/r_1) + a r & {\rm at}~~ r>r_c\,.
\end{array} \right.
\label{vsfhmn}
\eea
Here,
\bea
V_{\rm pert}(r) = - C_F
\frac{\alpha\rms{\overline{MS}}(\mu_2^2)}{r}
\left[
1 + a_1\frac{\alpha\rms{\overline{MS}}(\mu_2^2)}{4\pi}
+ \Big( a_2 + \frac{\pi^2 \beta_0^2}{3} \Big)
\Big(\frac{\alpha\rms{\overline{MS}}(\mu_2^2)}{4\pi}\Big)^2
\right]
\label{coorptpot}
\eea
represents the coordinate-space potential
in the second scheme,
$\mu_2 = \exp(-\gamma_E)/r$. 
The coefficients $a_1$ and $a_2$ are the same as in the momentum-space
potential, and
$\beta_0 = ( 11 C_A - 4 T_F n_f )/3$.
See Ref.~\cite{MP97} for details.\footnote{
We evolve the $\overline{\rm MS}$-coupling $\alpha\rms{\overline{MS}}(\mu)$
by solving the three-loop renormalization group equation {\it numerically}
for a given initial value at $\mu = M_Z$, whereas an approximate solution
to the renormalization group equation is used in~\cite{MP97}.
}

The values of the phenomenological parameters 
$r_0$, $r_1$, $a$ and $c_1$ are
taken from Ref.~\cite{sfhmn} and are tuned to reproduce bottomonium
and charmonium data well:
\bea
\begin{array}{llll}
      r_0 &=& 0.2350 & {\rm GeV}^{-1} \\
      r_1 &=& 3.745  & {\rm GeV}^{-1} \\
      a   &=& 0.3565 & {\rm GeV}^{-2} \\
      c_1 &=& 0.8789 & {\rm GeV}
\end{array}
\eea
We fix $c_0$ and $r_c$ by requiring that both the potential 
$V_{\rm SFHMN}(r)$ and
its first derivative are continuous at $r = r_c$.
For example, $r_c =  0.2526$~GeV$^{-1}$ and $c_0=-1.972$~GeV for 
$\alpha\rms{\overline{MS}}(M_Z^2)=0.115$.

\begin{figure}[htb]\begin{center}
 \epsfxsize 12cm \mbox{\epsfbox{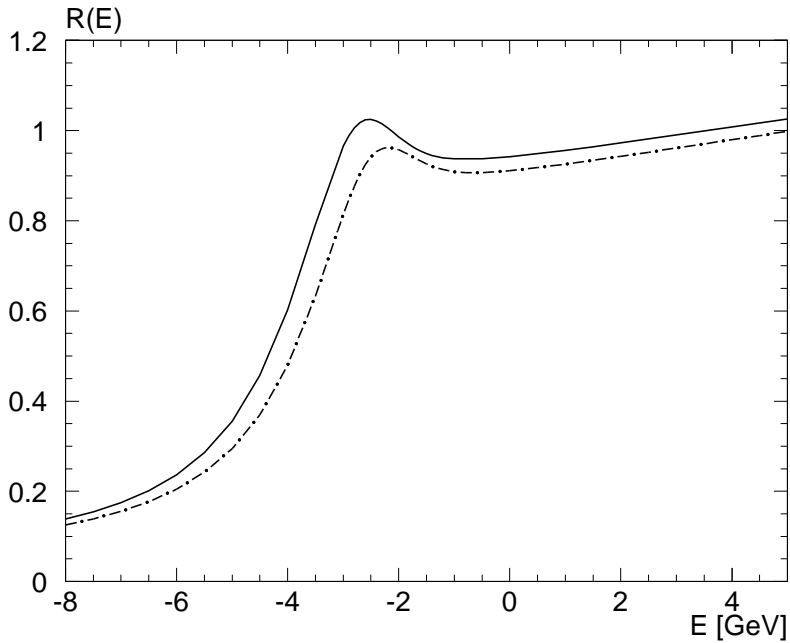}}
 \end{center}
 \caption[]{\label{compsfhmn}Comparison of the total cross section 
  (normalized to $R$) as a function of $E=\sqrt{s}-2m_t$ for $m_t=175$\ GeV,
  $\alpha\rms{\overline{MS}}=0.115$ and and $\Gamma_t = 1.427$~GeV 
  using the old and the present versions of the coordinate-space
  potential $V_{\rm SFHMN}$.
  The dash-dotted line
  corresponds to the potential described in~\cite{sfhmn}.
  The solid line shows the prediction of the present potential,
  Eq.~(\ref{vsfhmn}).}
\end{figure}

This potential is an improved version of the potential proposed
in~\cite{sfhmn} by including the next-to-next-to leading order
terms to the short-distance QCD potential.
We compare the cross sections calculated from the present version
and from the old version in Fig.~\ref{compsfhmn} for 
$\alpha\rms{\overline{MS}}(M_Z^2)=0.115$.

\end{appendix}

\newpage

\def\app#1#2#3{{\it Acta~Phys.~Polonica~}{\bf B #1} (#2) #3}
\def\apa#1#2#3{{\it Acta Physica Austriaca~}{\bf#1} (#2) #3}
\def\npb#1#2#3{{\it Nucl.~Phys.~}{\bf B #1} (#2) #3}
\def\plb#1#2#3{{\it Phys.~Lett.~}{\bf B #1} (#2) #3}
\def\prd#1#2#3{{\it Phys.~Rev.~}{\bf D #1} (#2) #3}
\def\pR#1#2#3{{\it Phys.~Rev.~}{\bf #1} (#2) #3}
\def\prl#1#2#3{{\it Phys.~Rev.~Lett.~}{\bf #1} (#2) #3}
\def\sovnp#1#2#3{{\it Sov.~J.~Nucl.~Phys.~}{\bf #1} (#2) #3}
\def\yadfiz#1#2#3{{\it Yad.~Fiz.~}{\bf #1} (#2) #3}
\def\jetp#1#2#3{{\it JETP~Lett.~}{\bf #1} (#2) #3}
\def\zpc#1#2#3{{\it Z.~Phys.~}{\bf C #1} (#2) #3}

\end{document}